\documentclass{emulateapj}
\usepackage[colorlinks,citecolor=blue]{hyperref} 

\def\degrees{^\circ}
\def\kms{{\ }{\rm km}\,{\rm s}^{-1}}

\begin{document}

\submitted{Accepted by ApJL on January 21, 2017}
           
\slugcomment{{\it Accepted by ApJL on January 21, 2017}}
                  
\shortauthors{KAZANTZIDIS ET AL.}

\shorttitle{The Effects of Ram-pressure Stripping and Supernova Winds
  on the Tidal Stirring of Disky Dwarfs}

\title{The Effects of Ram-pressure Stripping and Supernova Winds on the Tidal\\ 
  Stirring of Disky Dwarfs: Enhanced Transformation into Dwarf Spheroidals}

\author{Stelios Kazantzidis\altaffilmark{1},  
        Lucio Mayer\altaffilmark{2},
        Simone Callegari\altaffilmark{3},\\ 
        Massimo Dotti\altaffilmark{4},
        and Leonidas A. Moustakas\altaffilmark{5}}

\altaffiltext{1}{Section of Astrophysics, Astronomy and Mechanics,
  Department of Physics, National and Kapodistrian University of
  Athens, 15784 Zografos, Athens, Greece; skazantzidis@phys.uoa.gr}

\altaffiltext{2}{Center for Theoretical Astrophysics and Cosmology,
  Institute for Computational Science, University of Z\"urich, CH-8057
  Z\"urich, Switzerland}

\altaffiltext{3}{Anthropology Institute and Museum, University of
  Z\"urich, CH-8057 Z\"urich, Switzerland}

\altaffiltext{4}{Universit\`a degli Studi di Milano-Bicocca, Piazza
  della Scienza 3, I-20126 Milano, Italy}

\altaffiltext{5}{Jet Propulsion Laboratory, California Institute of
  Technology, Pasadena, CA 91109, USA}

\begin{abstract}

A conclusive model for the formation of dwarf spheroidal (dSph)
galaxies still remains elusive. Owing to their proximity to the
massive spirals Milky Way (MW) and M31, various environmental
processes have been invoked to explain their origin. In this context,
the tidal stirring model postulates that interactions with MW-sized
hosts can transform rotationally supported dwarfs, resembling
present-day dwarf irregular (dIrr) galaxies, into systems with the
kinematic and structural properties of dSphs. Using $N$-body$+$SPH
simulations, we investigate the dependence of this transformation
mechanism on the gas fraction, $f_{\rm gas}$, in the disk of the
progenitor dwarf. Our numerical experiments incorporate for the first
time the combined effects of radiative cooling, ram-pressure
stripping, star formation, supernova (SN) winds, and a cosmic UV
background. For a given orbit inside the primary galaxy, rotationally
supported dwarfs with gas fractions akin to those of observed dIrrs
($f_{\rm gas} \gtrsim 0.5$), demonstrate a substantially enhanced
likelihood and efficiency of transformation into dSphs relative to
their collisionless ($f_{\rm gas} =0$) counterparts. We argue that the
combination of ram-pressure stripping and SN winds causes the gas-rich
dwarfs to respond more impulsively to tides, augmenting their
transformation. When $f_{\rm gas} \gtrsim 0.5$, disky dwarfs on
previously unfavorable low-eccentricity or large-pericenter orbits are
still able to transform. On the widest orbits, the transformation is
incomplete; the dwarfs retain significant rotational support, a
relatively flat shape, and some gas, naturally resembling
transition-type systems. We conclude that tidal stirring constitutes a
prevalent evolutionary mechanism for shaping the structure of dwarf
galaxies within the currently favored CDM cosmological paradigm.

\end{abstract}


\section{Introduction}
\label{sec:introduction}

The Local Group (LG) and its population of dwarf spheroidal (dSph)
galaxies provide crucial insights into cosmic structure
formation. These extraordinary objects are the faintest and most dark
matter (DM) dominated galaxies known
\citep[e.g.,][]{Mateo98,McConnachie12}. Furthermore, dSphs are gas
poor or completely devoid of gas \citep[e.g.,][]{Grcevich_Putman09}
and they are characterized by pressure-supported, spheroidal stellar
components \citep[e.g.,][]{Mateo98} and widely diverse star formation
(SF) histories \citep[e.g.,][]{Skillman_etal17}.

Given that dSphs are preferentially located near the massive spirals
Milky Way (MW) and M31, various environmental mechanisms have been
invoked to explain their origin
\citep[e.g.,][]{Einasto_etal74,Faber_Lin83,Kroupa97,Mayer_etal01a,Mayer_etal01b,
  Mayer_etal06,Mayer_etal07,Kravtsov_etal04,D'Onghia_etal09,Klimentowski_etal09,
  Kazantzidis_etal11,Kazantzidis_etal13,Pawlowski_etal11,Yozin_Bekki12}\footnote{For
  alternative scenarios of dSph formation, see, e.g.,
  \citet{Sawala_etal10} and \citet{Assmann_etal13}.}. In this context,
the ``tidal stirring'' model \citep{Mayer_etal01a} posits that
gravitational and hydrodynamical interactions with MW-sized hosts can
transform rotationally supported dwarfs, resembling present-day dwarf
irregulars (dIrrs), into systems with the kinematic and structural
properties of dSphs.

Our study expands upon all previous investigations of tidal stirring
in one major respect; the inclusion, for the first time, of the
combined effects of ram-pressure stripping and supernova (SN) winds
driving gas bulk motions and outflows. Such winds can directly affect
the mass distribution of dwarf galaxies, significantly decreasing
their total central density
\citep[e.g.,][]{Governato_etal10,Teyssier_etal13,Shen_etal14,Onorbe_etal15,Read_etal16}. In
fact, any physical process (e.g., ram-pressure stripping) that can
cause the removal of gas in dwarfs may reduce the inner density of
their DM halos
\citep[e.g.,][]{Navarro_etal96a,Read_Gilmore05,Arraki_etal14}.

This improvement in the modeling may be vital to tidal
stirring. Indeed, systems with lower central densities are
characterized by longer internal dynamical times and are thus expected
to respond more impulsively to external tidal perturbations and suffer
stronger tidal shocks \citep[e.g.,][]{Gnedin_Ostriker99}. Therefore,
rotationally supported dwarfs with decreased central densities due to
baryonic effects may experience increasingly probable and efficient
transformations into dSphs. 

Here we explore this qualitative expectation via a series of tidal
stirring simulations of disky dwarfs with varying fractions of their
disk mass in gas. Our results indicate that the likelihood and
efficiency of transformation into a dSph are both enhanced
significantly when the progenitor rotationally-supported dwarfs are
gas-rich, with gas fractions that are in accordance with those
inferred from observations and recent theoretical efforts in dwarf
galaxy formation. This finding further establishes tidal stirring as a
prevalent evolutionary mechanism for shaping the structure of dwarf
galaxies within the currently favored CDM cosmological paradigm.

\section{Methods}
\label{sec:methods}

We employed the technique of \citet{Widrow_etal08} to create
self-consistent N-body realizations of rotationally supported dwarf
galaxies consisting of exponential, baryonic disks embedded in cuspy,
cosmologically-motivated \citet{Navarro_etal96b} DM halos.

Our goal is to elucidate the degree to which dSph formation via tidal
stirring is affected by the presence of a gaseous component in the
disk of the progenitor rotationally supported dwarf. To accomplish
this, we varied the fraction of the disk mass in gas, $f_{\rm gas}$,
in three otherwise identically initialized disky dwarf galaxies;
$f_{\rm gas}=0$, $f_{\rm gas}=0.5$, and $f_{\rm gas}=0.8$. The adopted
values of $f_{\rm gas}=0.5$ are well-motivated as they are akin to
those of both observed dIrrs \citep[e.g.,][]{McConnachie12,Oh_etal15}
and simulated, realistic dIrr-like systems
\citep[e.g.,][]{Governato_etal10,Shen_etal14}.

Following \citet{Widrow_etal08}, we first generated the dwarf model
with a purely stellar disk ($f_{\rm gas}=0$). Subsequently, we
converted a fraction $f_{\rm gas}$ of randomly selected stellar
particles in the dwarf disk into gas particles, without changing their
positions and masses \citet{Dotti_etal07}. Gas particles were set on
tangential orbits, with the dynamical equilibrium achieved by
enforcing vertical pressure support to the disk, and by correcting
their tangential velocities for the radial component of the pressure
gradient force.

Each dwarf comprised a DM halo with a virial mass of $M_{\rm vir} =
10^{10} M_{\odot}$, a choice that is motivated by theoretical studies
of the mass distribution of satellites that could correspond to
present-day dSphs
\citep[e.g.,][]{Kravtsov_etal04,Klimentowski_etal10,Tomozeiu_etal16}. Informed
by the same investigations, we constructed our dwarfs considering a
$z=1$ infall redshift onto the primary. The virial radius and the
median concentration value for a $z=1$ cosmological halo at this mass
scale are $r_{\rm vir} \approx 32.9$~kpc and $c \approx 9.4$,
respectively \citep[e.g.,][]{Maccio_etal07}.

All dwarf galaxies hosted an identical, exponential baryonic disk
whose mass, sech$^2$ vertical scale-height, and central radial
velocity dispersion were $M_d=0.02 M_{\rm vir}$, $z_d =0.2 R_d$, and
$\sigma_{R0}=15\kms$, respectively
\citep{Kazantzidis_etal11}. Assuming a dimensionless halo spin
parameter of $\lambda=0.04$ yields a disk radial scale-length of $R_d
\approx 0.76$~kpc \citep{Mo_etal98}.

Numerical parameters (i.e., particle numbers and gravitational
softening lengths) for each component of the dwarf models are listed
in Table~\ref{table:numerical_parameters}. For a given component,
resolution (mass and force) was constant. This is necessary in order
to perform a numerically robust comparison among different
simulations. We checked the adequacy of our dwarf models by evolving
them for several Gyr in isolation (using an adiabatic equation of
state for the gas-rich dwarfs) and confirming that they retained their
equilibrium configuration.

For simplicity, we assumed a single host represented by a
self-gravitating, high-resolution MW model
\citep{Kazantzidis_etal11}. To explore the effect of ram pressure, we
included an extended, non-rotating halo of hot gas in hydrostatic
equilibrium inside the primary galaxy \citep{Mastropietro_etal05} and
with properties consistent with MW observations
\citep{Miller_Bregman13}. We modeled this component with $6.25 \times
10^6$ particles and employed a fairly large gravitational softening,
$\epsilon=2$~kpc, to minimize discreteness noise in the host
potential. The mass ratio of the gas particles in the dwarfs and those
in the hot gas halo was $1$, preventing spurious enhancement of
ram-pressure stripping \citep{Abadi_etal99}.

Given that the $f_{\rm gas}=0.8$ disky dwarf is characterized by the
largest gas fraction, we use it as the basis for the comparison with
the $f_{\rm gas}=0$ model. To this end, each of these dwarfs was
placed on seven bound orbits (O1--O7) of varying sizes and
eccentricities inside the primary
galaxy. Table~\ref{table:orbital_parameters} summarizes the adopted
orbital parameters, which are motivated by theoretical studies of the
orbital distributions of cosmological satellites in MW-sized hosts
\citep[e.g.,][]{Diemand_etal07,Klimentowski_etal10}.

For completeness, we performed $8$ additional simulations, for a total
of $22$ experiments (S1--S22;
Table~\ref{table:summary}). Specifically, to ascertain the relative
effect of $f_{\rm gas}$ on dSph formation, we followed the evolution
of dwarf model $f_{\rm gas}=0.5$ on orbits O1--O3
(S15--S17). Moreover, in an effort to investigate whether the cosmic
epoch at which the disky dwarfs were accreted by their hosts could
affect their transformation, we evolved model $f_{\rm gas}=0.8$ on
orbits O1--O3, assuming a larger $z=2$ infall redshift onto the
primary (S18--S20). This is relevant as a stronger cosmic UV
background is expected to augment rampressure stripping by enhancing
gas heating and ionization. In the latter set of simulations, we only
modified the time evolution of the intensity of the cosmic UV
background to reflect a $z=2$ accretion epoch. To avoid introducing an
additional parameter, the internal structure of the disky dwarf was
explicitly not altered in these cases. We note that in all experiments
described above (S1--S20), the alignments between the internal angular
momenta of the dwarfs, that of the primary disk, and that of the
orbital angular momenta, were all {\it mildly} prograde and equal to
$45\degrees$ \citep{Kazantzidis_etal11}.

Lastly, we explored the importance of the initial inclination of the
dwarf disk with respect to the orbital plane by placing model $f_{\rm
  gas}=0.8$ on orbit O1 and changing the default value from
$i=45\degrees$ to $i=0\degrees$ and to $i=90\degrees$ (S21--S22).
These two configurations correspond to a perfectly edge-on and a
perfectly face-on disk and bracket the range of possible amplitudes
for the ram-pressure force.

\section{Simulations}
\label{sec:simulations}

All simulations were performed with the parallel Tree$+$SPH code {\it
  GASOLINE} \citep{Wadsley_etal04}. We included Compton cooling,
atomic cooling, and metallicity-dependent radiative cooling at low
temperatures \citep{Mashchenko_etal06}. A uniform, time-variable
cosmic UV background, causing the photoionization and photoheating of
the gas, was implemented using the \citet{Haardt_Madau12} model. We
incorporated the effects of SF and SN feedback (``blastwave'' scheme)
according to \citet{Stinson_etal06}. These recipes are characterized
by several parameters, including the gas density threshold for SF,
$n_{\rm SF}$, and we adopted the relevant values from
\citet{Governato_etal10}. We emphasize that our high SF density
threshold ($n_{\rm SF}=10$ atoms cm$^{-3}$) enables the development of
a realistic clumpy, inhomogeneous ISM, where SF and energy injection
from SN explosions occur in a clustered fashion. As we demonstrate
below, this has profound consequences on dSph formation via tidal
stirring. Lastly, our gas mass resolution and the employed value of
$n_{\rm SF}$ ensure that there is no artificial fragmentation
\citep{Bate_Burkert97}.


\begin{table}
\caption{Numerical Parameters of Dwarf Models}
\begin{center}
  \vspace*{-12pt}
\begin{tabular}{lccccccc}
\hline
\hline 
\vspace*{-8pt}
\\
\multicolumn{1}{c}{Model}                      & 
\multicolumn{1}{c}{$f_{\rm gas}$}                &
\multicolumn{1}{c}{$N_{\rm DM}$}                 &
\multicolumn{1}{c}{$N_{\ast}$}                   &
\multicolumn{1}{c}{$N_{\rm gas}$}                & 
\multicolumn{1}{c}{$\epsilon_{\rm DM}$}          &
\multicolumn{1}{c}{$\epsilon_{\ast}$}            &
\multicolumn{1}{c}{$\epsilon_{\rm gas}$}       
\\
\multicolumn{1}{c}{}                           & 
\multicolumn{1}{c}{}                           & 
\multicolumn{1}{c}{}                           & 
\multicolumn{1}{c}{}                           & 
\multicolumn{1}{c}{}                           & 
\multicolumn{1}{c}{(pc)}                       & 
\multicolumn{1}{c}{(pc)}                       &
\multicolumn{1}{c}{(pc)}                    
\vspace*{-8pt}
\\
\\
\hline
\vspace*{-8pt}
\\
D1 &  0    & $10^6$ & $2 \times 10^6$   & ...                 & 100 & 15 & ... \\
D2 &  0.5  & $10^6$ & $10^6$            & $6.25 \times 10^4$  & 100 & 15 & 50  \\
D3 &  0.8  & $10^6$ & $4 \times 10^5$   & $10^5$              & 100 & 15 & 50  \\
\vspace*{-7pt}
\\
\hline
\end{tabular}
\end{center}
\vspace{-0.2cm}
\label{table:numerical_parameters}
\end{table}



\begin{table}
\caption{Orbital Parameters of Disky Dwarfs}
\begin{center}
  \vspace*{-12pt}
\begin{tabular}{lccc}
\hline
\hline 
\vspace*{-8pt}
\\
\multicolumn{1}{c}{Orbit}                      & 
\multicolumn{1}{c}{$r_{\rm apo}$}                &
\multicolumn{1}{c}{$r_{\rm peri}$}               &
\multicolumn{1}{c}{$r_{\rm apo}/r_{\rm peri}$}   
\\
\multicolumn{1}{c}{}                           & 
\multicolumn{1}{c}{(kpc)}                      & 
\multicolumn{1}{c}{(kpc)}                      &
\multicolumn{1}{c}{}                           
\vspace*{-8pt}
\\
\\
\hline
\vspace*{-8pt}
\\
O1 &  125  &  25    &  5    \\
O2 &  85   &  17    &  5    \\
O3 &  250  &  50    &  5    \\

O4 &  125  &  12.5  &  10   \\
O5 &  125  &  50    &  2.5  \\
O6 &  80   &  50    &  1.6  \\
O7 &  250  &  12.5  &  20   \\
\vspace*{-7pt}
\\
\hline
\end{tabular}
\end{center}
\vspace{-0.2cm}
\label{table:orbital_parameters}
\end{table}



\begin{figure*}[t]
\begin{center}
\begin{tabular}{c}
  \includegraphics[scale=0.42]{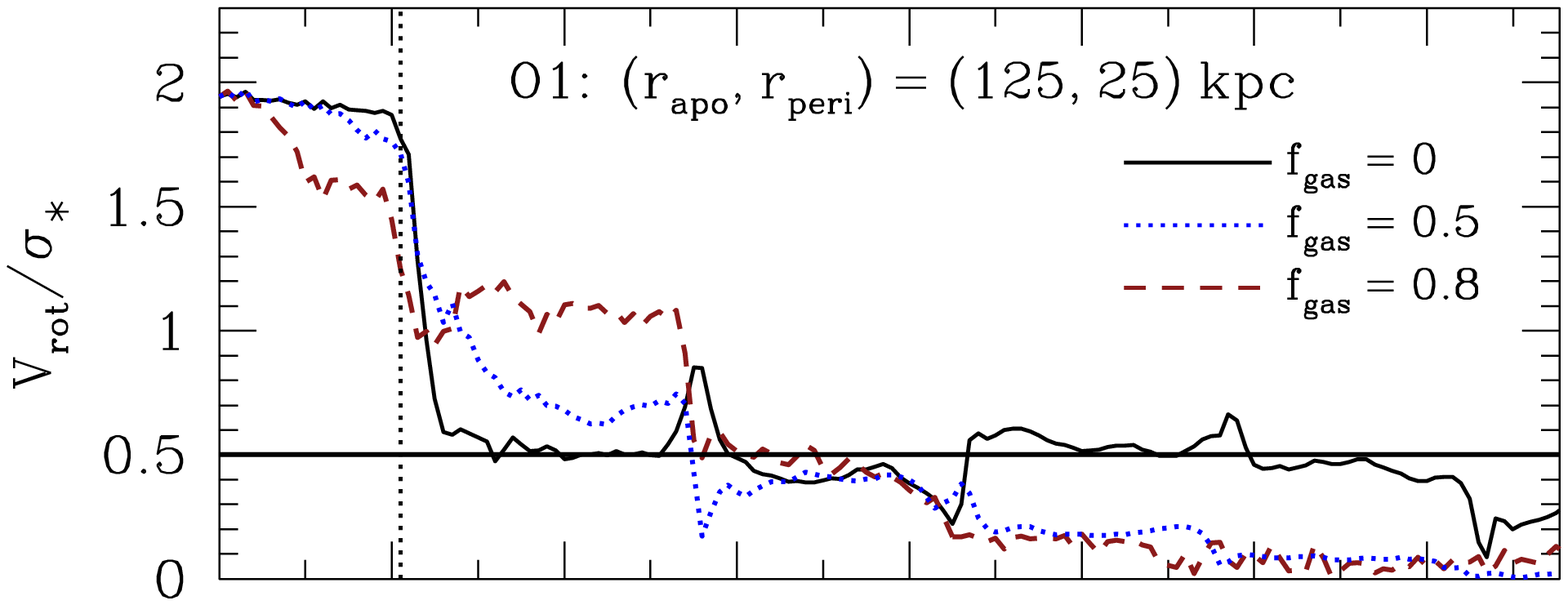}\\
  \includegraphics[scale=0.42]{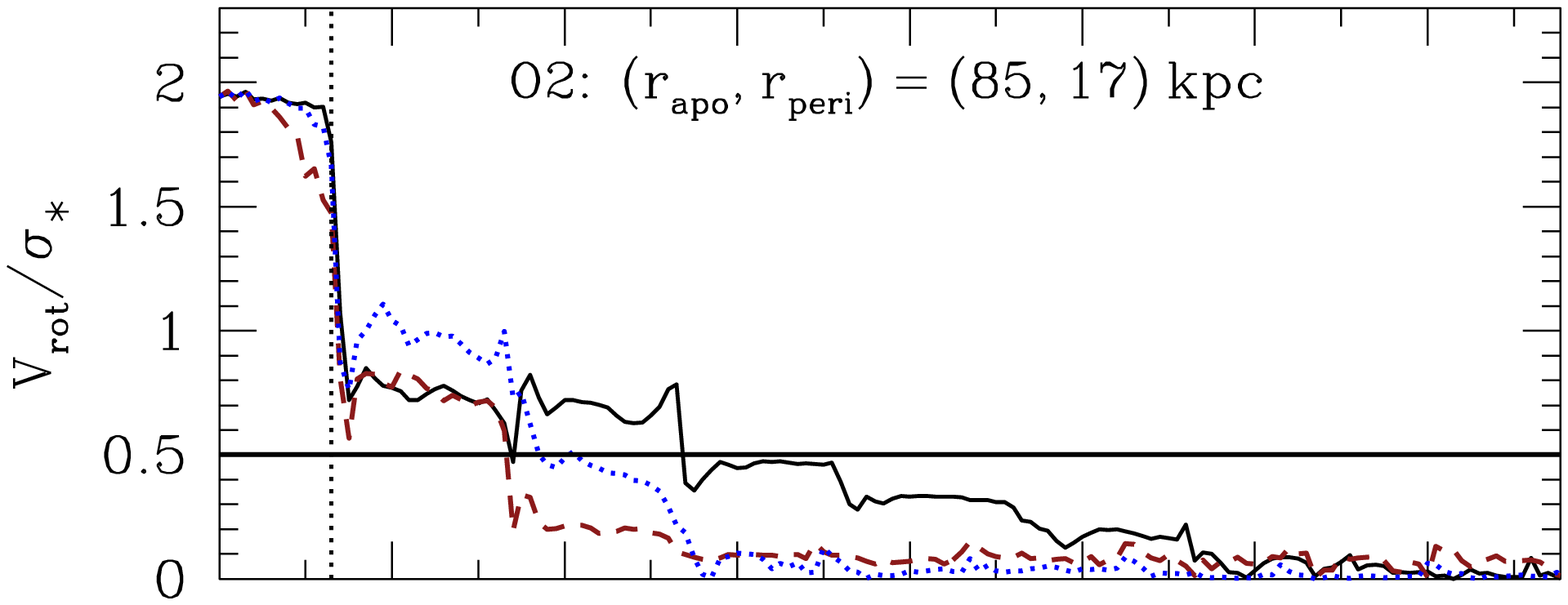}\\
  \includegraphics[scale=0.42]{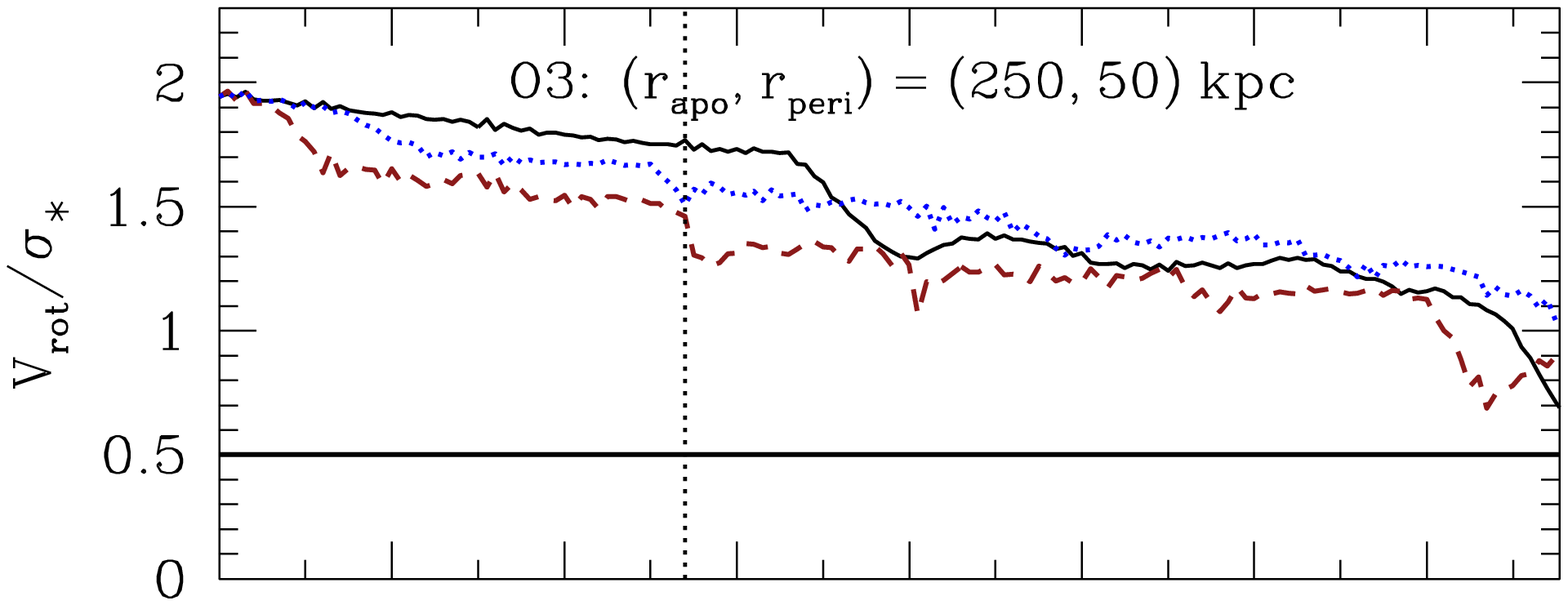}\\
  \includegraphics[scale=0.42]{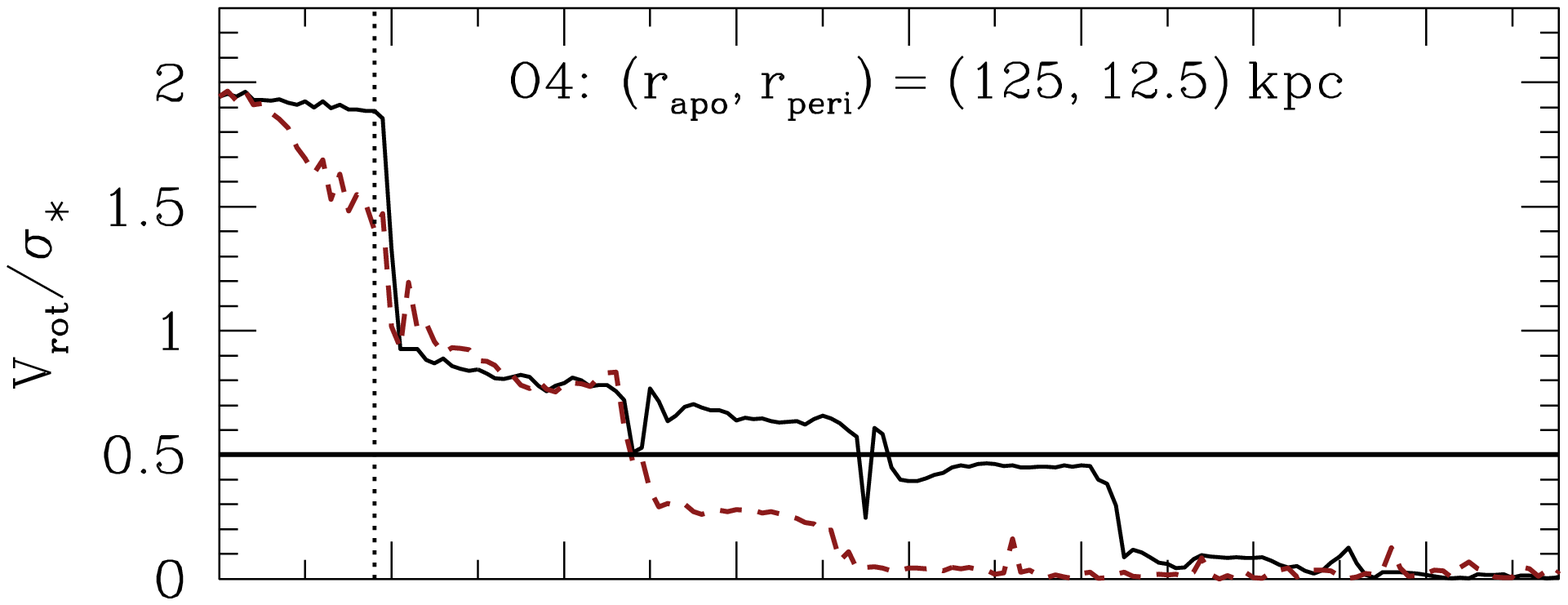}\\
  \includegraphics[scale=0.42]{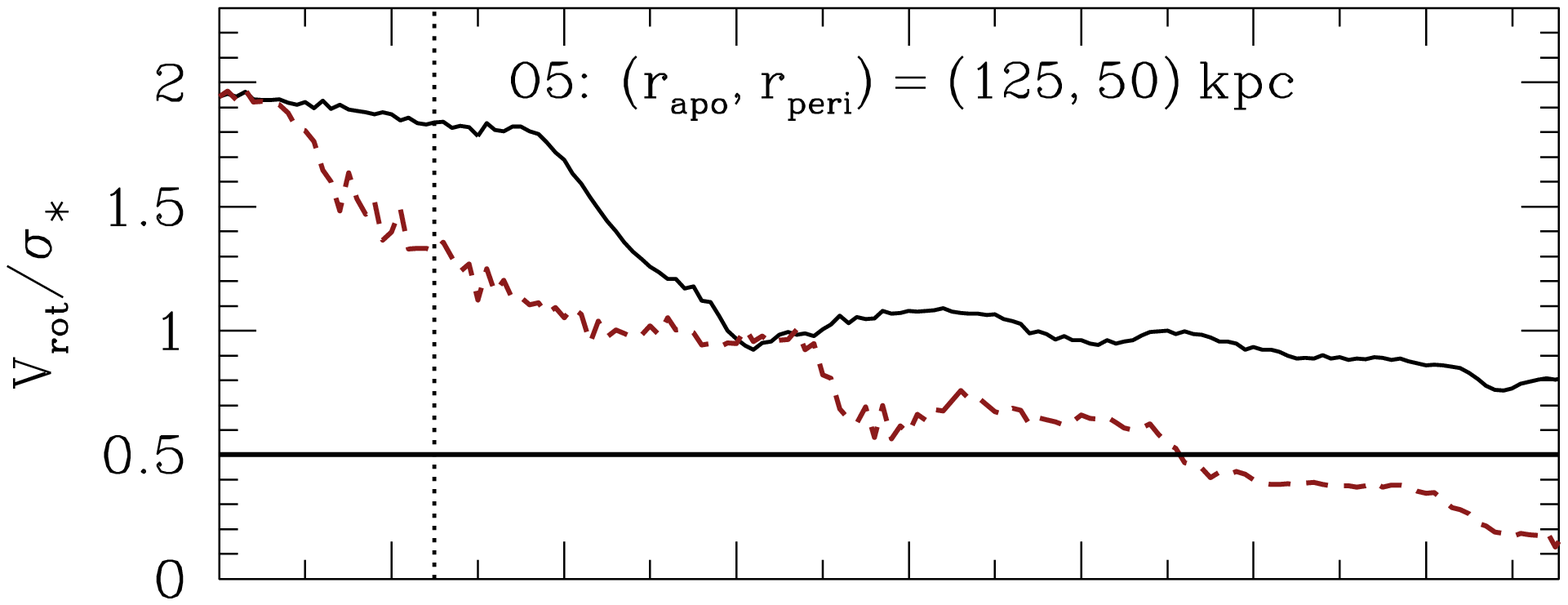}\\
  \includegraphics[scale=0.42]{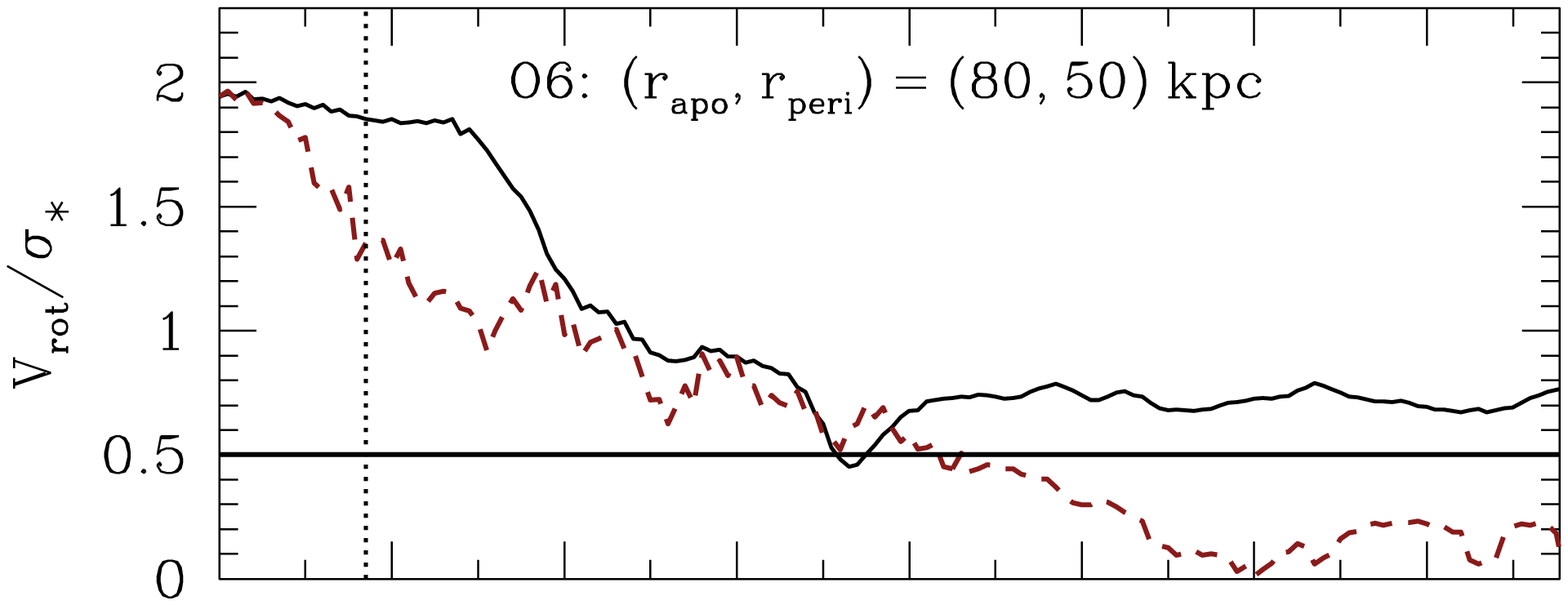}\\
  \includegraphics[scale=0.42]{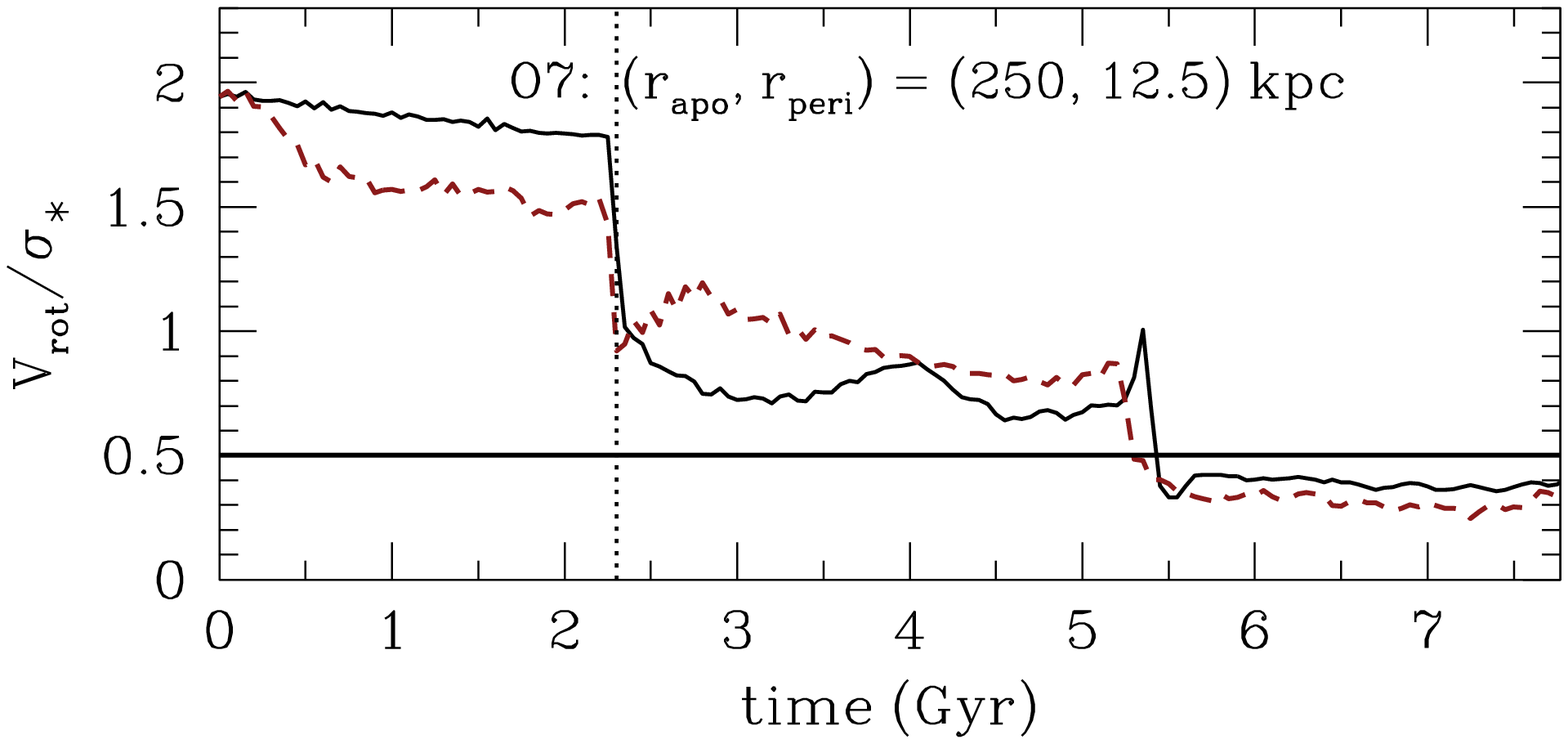}
\end{tabular}
\begin{tabular}{c}
  \includegraphics[scale=0.42]{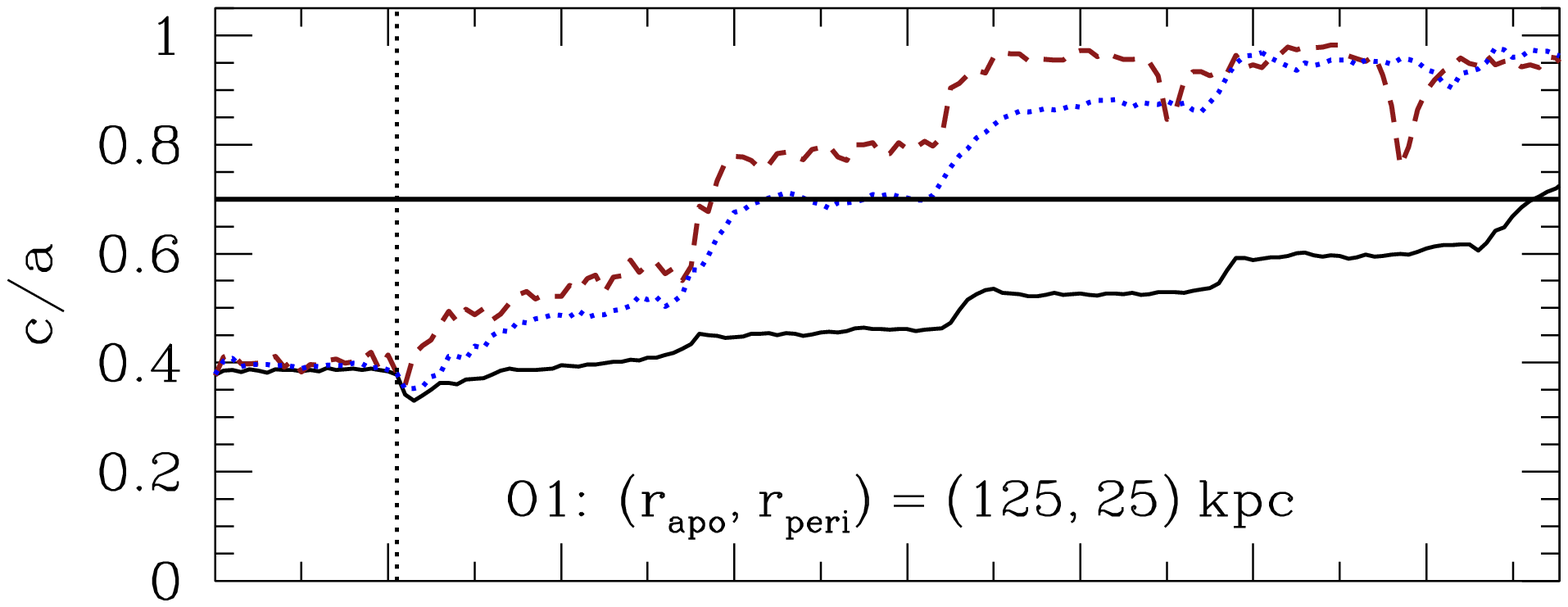}\\
  \includegraphics[scale=0.42]{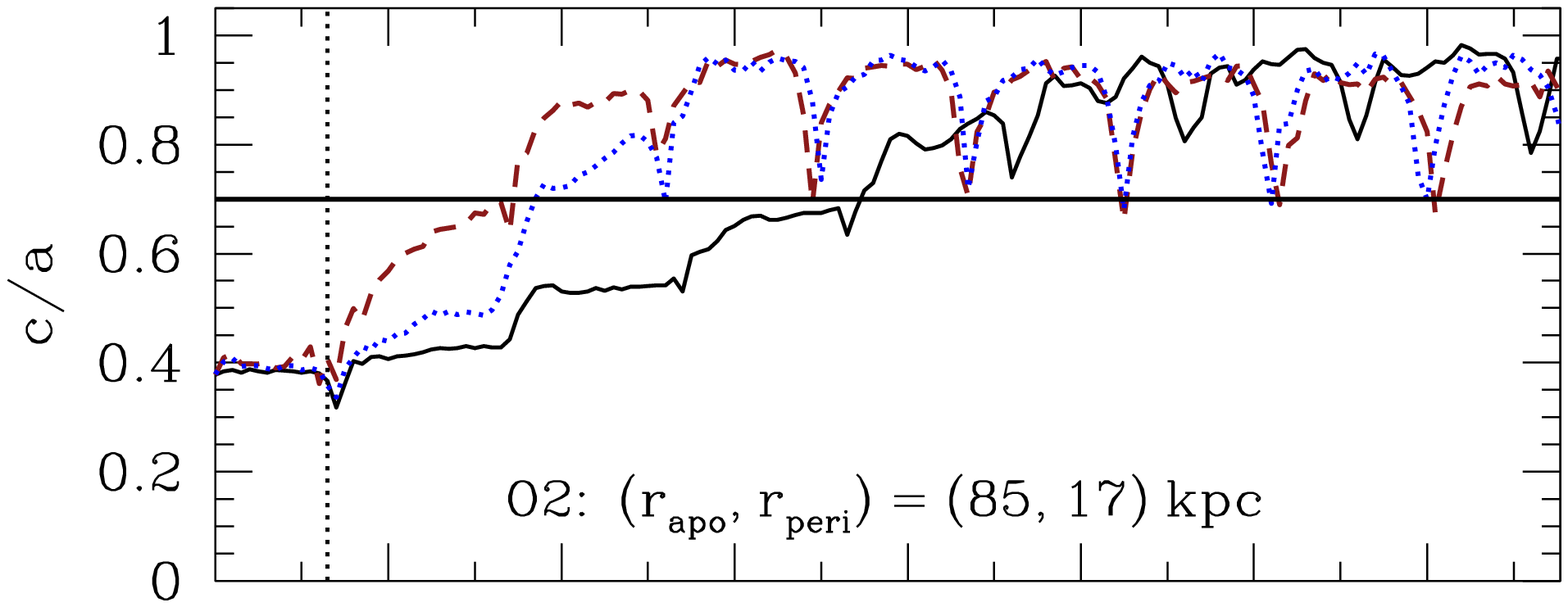}\\
  \includegraphics[scale=0.42]{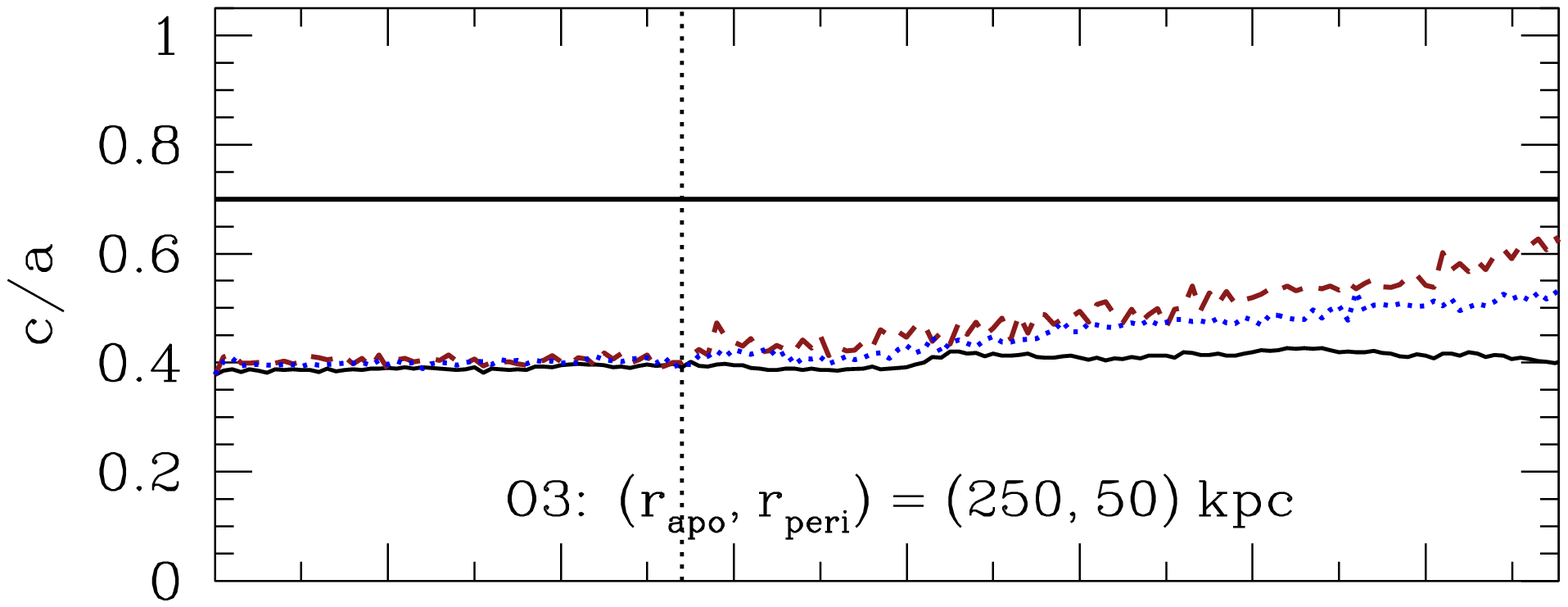}\\
  \includegraphics[scale=0.42]{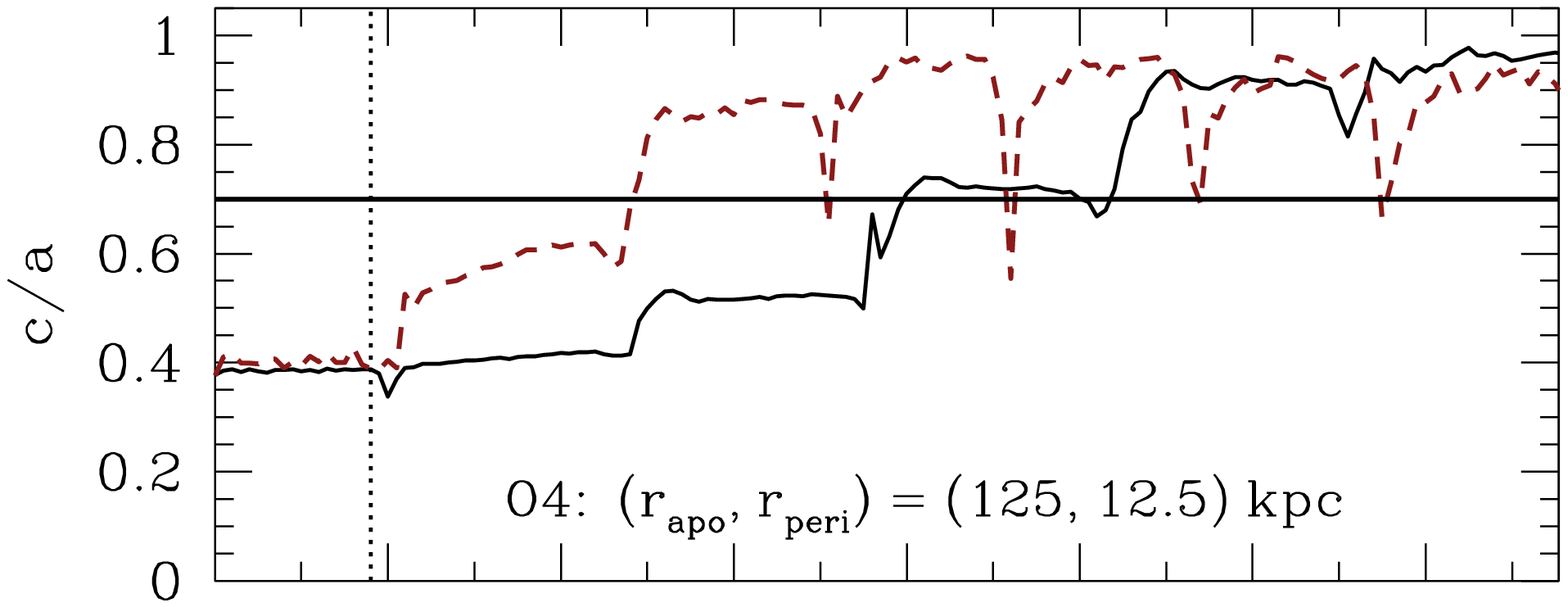}\\
  \includegraphics[scale=0.42]{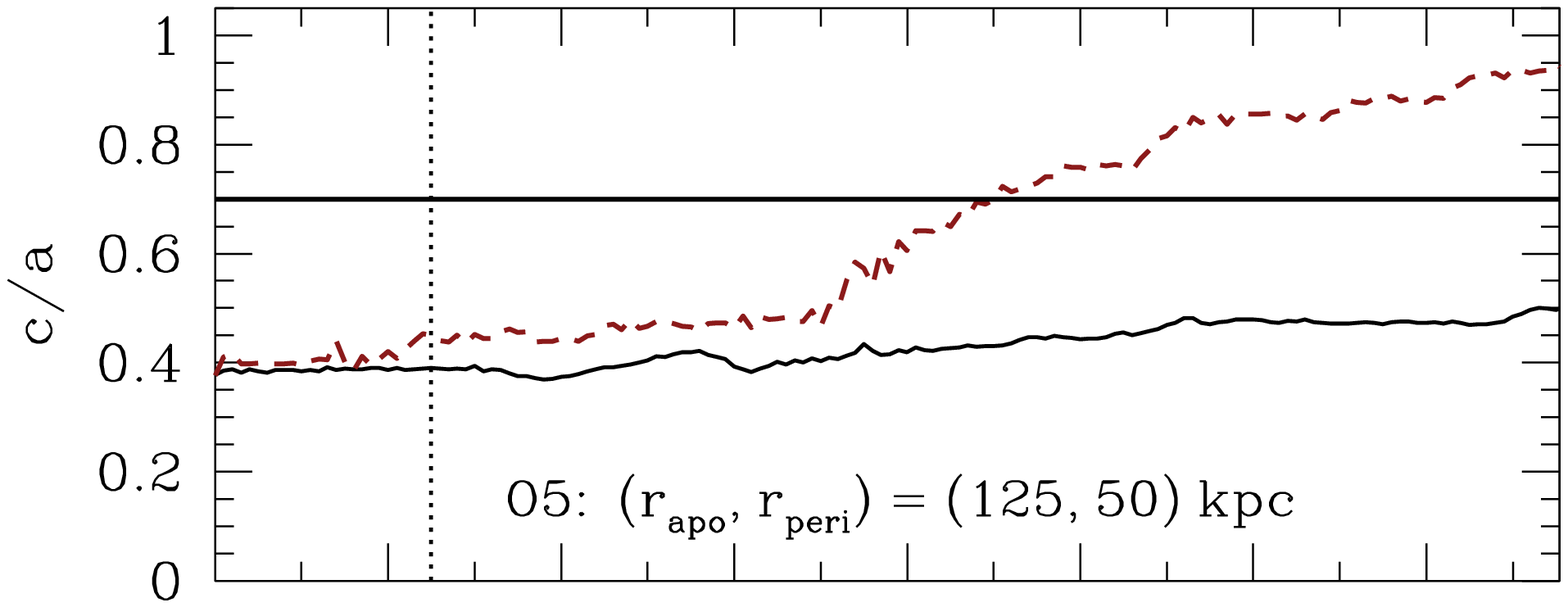}\\
  \includegraphics[scale=0.42]{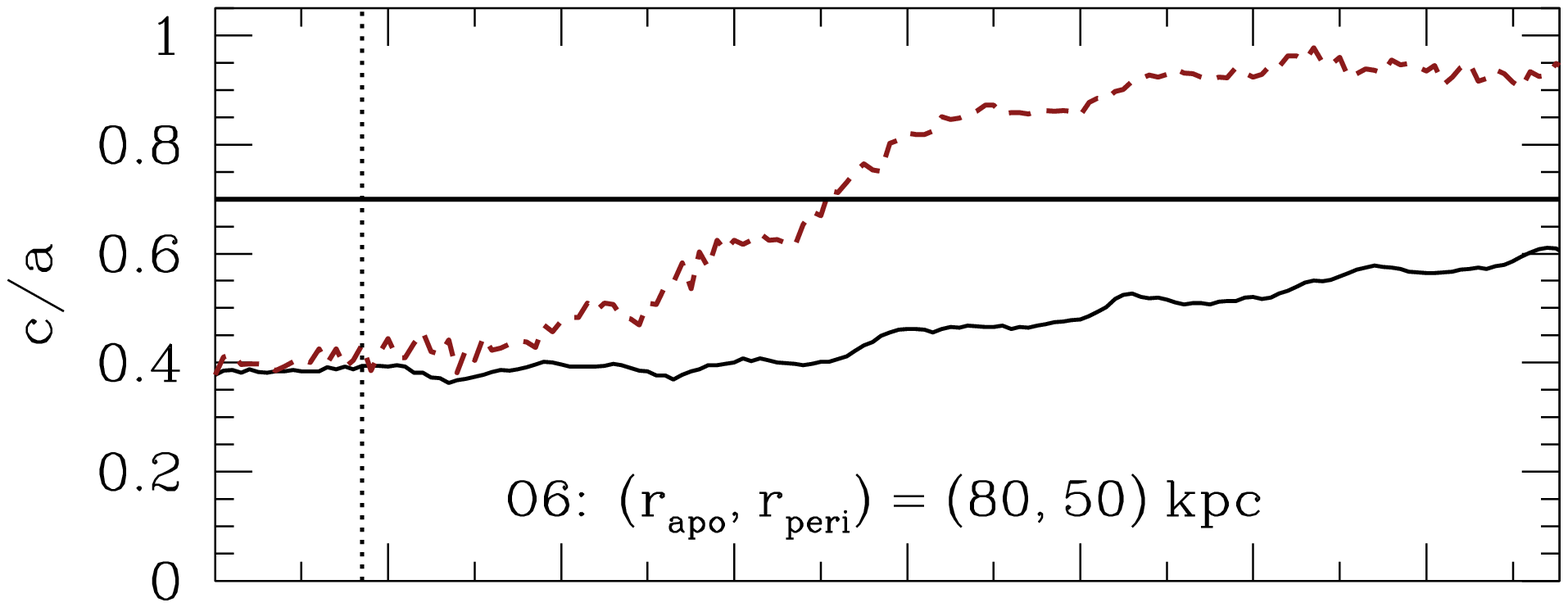}\\
  \includegraphics[scale=0.42]{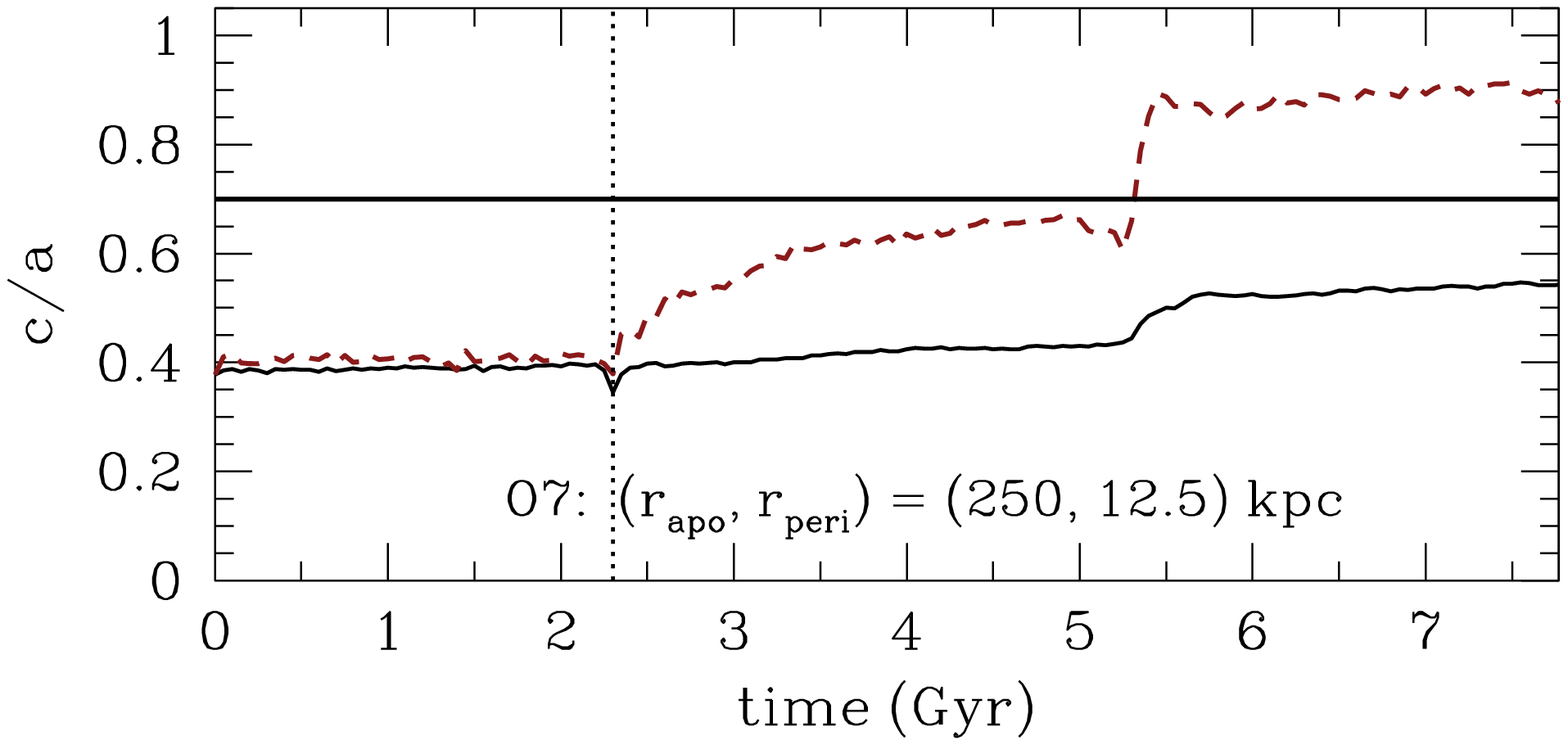}
\end{tabular}
\end{center}
\vspace{-0.5cm}
\caption{Evolution of the stellar kinematics (left panels) and shapes
  (right panels) of the simulated disky dwarfs as a function of
  time. The results are presented for orbits O1--O7. $f_{\rm gas}$
  denotes the fraction of the disk mass in gas and the solid, dashed,
  and dotted lines correspond to $f_{\rm gas}=0$, $f_{\rm gas}=0.5$,
  and $f_{\rm gas}=0.8$, respectively. In each panel, the vertical
  line specifies the initial pericentric passage. The horizontal lines
  indicate the limiting values $V_{\rm rot}/\sigma_{\ast} =0.5$, and
  $c/a=0.7$: simulated dwarf galaxies whose final states are
  characterized by $V_{\rm rot}/\sigma_{\ast} \lesssim 0.5$, $c/a
  \gtrsim 0.7$, and $M_{\rm gas} \simeq 0$ correspond to bona fide
  dSphs. For a given orbit, gas-rich ($f_{\rm gas} \gtrsim 0.5$)
  rotationally supported dwarfs experience a stronger evolution in
  their shapes and kinematics and exhibit a considerably enhanced
  likelihood and efficiency of transformation into dSph-like systems
  relative to their collisionless ($f_{\rm gas} =0$) counterparts.
  \label{fig1}}
\end{figure*}


\section{Results}
\label{sec:results}

The response of the rotationally supported dwarfs to the host galaxy
tidal field is assessed through the evolution of their masses,
kinematics, and shapes. These properties were always computed within
$0.76$~kpc, a radius that corresponds to the scale-length of the
initial disk. Employing this well-defined, fixed scale allows us to
overcome the complications related to determining tidal radii
\citep[e.g.,][]{Read_etal06} and aids the comparison among different
experiments.

We used the parameters $c/a$ and $V_{\rm rot}/\sigma_{\ast}$ c/a to
quantify the shapes and kinematics of the dwarf galaxies: $c$, $a$,
$V_{\rm rot}$, and $\sigma_{\ast}$ correspond to the minor axis, the
major axis, the rotational velocity, and the one-dimensional velocity
dispersion of the stellar distribution, respectively. At each
simulation output, we determined the directions of the principal axes
and computed $c/a$ via the moments of the inertia
tensor. Subsequently, we calculated the rotational velocity around the
minor axis, $V_{\rm rot}=V_{\phi}$, and the dispersions $\sigma_r$,
$\sigma_{\theta}$, and $\sigma_{\phi}$ around the mean values;
$\sigma_{\ast}$, which measures the amount of stellar random motions,
is defined as $\sigma_{\ast} \equiv [(\sigma_r^2 + \sigma_{\theta}^2 +
  \sigma_{\phi}^2)/3]^{1/2}$.

Figure~\ref{fig1} shows the time evolution of the kinematics and
shapes of the simulated disky dwarfs as they orbit inside the
primary. In agreement with previous studies
\citep[e.g.,][]{Kazantzidis_etal11,Kazantzidis_etal13}, the repeated
action of the host galaxy tidal field causes simultaneously a
progressive decrease of $V_{\rm rot}/\sigma_{\ast}$ and a continuous
increase of $c/a$. Such evolution, which is strongly associated with
pericentric passages where tidal shocks occur, designates the gradual
transformation of the disky dwarfs into spheroidal stellar systems
dominated by random motions.

Our goal is to determine the likelihood and efficiency of dSph
formation via tidal stirring. To this end, we classify as bona fide
dSphs only those simulated dwarfs whose final states are characterized
by $V_{\rm rot}/\sigma_{\ast} \lesssim 0.5$, $c/a \gtrsim 0.7$, and
$M_{\rm gas} \simeq 0$ \citep[e.g.,][]{Mateo98, McConnachie12}.  We
underscore that the criterion $V_{\rm rot}/\sigma_{\ast} \lesssim 0.5$
is fairly conservative \citep{Kazantzidis_etal13}. Indeed, we measure
$V_{\rm rot}$ around the minor axis of the stellar distribution, which
corresponds to observing the simulated dwarfs perfectly
edge-on. Adopting a random line-of-sight would result in smaller
$V_{\rm rot}/\sigma_{\ast}$ values, indicating even more complete
transformations.

Table~\ref{table:summary} summarizes our results. During the orbital
evolution of the dwarfs, the strong tidal forces at pericenters may
trigger bar instabilities in their disks; column 6 refers to whether
such a tidally induced bar was formed\footnote{Following
  \citet{Kazantzidis_etal11}, we designate bar formation when the
  amplitude of the $m = 2$ Fourier component satisfies $A_2 \gtrsim
  0.2$ between two consecutive pericentric passages.}. Columns 7--9
list the final values of $M_{\rm gas}$, $V_{\rm rot}/\sigma_{\ast}$
and $c/a$, respectively. Column 10 specifies whether a dSph was
produced according to our criteria. Column 11 reports the time elapsed
from the beginning of the simulation until dSph formation occurs (the
number of corresponding pericentric passages is included in
parentheses).

Of the collisionless disky dwarfs, only those on high-eccentricity
($r_{\rm apo}/r_{\rm peri} \gtrsim 5$) and small-pericenter ($r_{\rm
  peri} \lesssim 25$~kpc) orbits that have also experienced at least
three pericentric passages are transformed into dSphs. Interestingly,
the likelihood and efficiency of dSph formation are both enhanced
significantly for gas-rich ($f_{\rm gas} \gtrsim 0.5$) dwarfs; in such
cases, not only the transformation into a dSph occurs on previously
unfavorable, low-eccentricity and/or large-pericenter orbits, but also
the number of required pericentric passages for dSph formation is
invariably smaller.


\begin{table*}
\caption{Summary of Results}
\begin{center}
  \vspace*{-12pt}
\begin{tabular}{lccccccccccc}
\hline
\hline
\vspace*{-8pt}
\\
\multicolumn{1}{c}{Simulation}                 &
\multicolumn{1}{c}{$f_{\rm gas}$}                & 
\multicolumn{1}{c}{$z$}                        &
\multicolumn{1}{c}{$i$}                        &
\multicolumn{1}{c}{Orbit}                      &
\multicolumn{1}{c}{Bar}                        &
\multicolumn{1}{c}{$M_{\rm gas}$}                &
\multicolumn{1}{c}{$V_{\rm rot}/\sigma_{\ast}$}   & 
\multicolumn{1}{c}{$c/a$}                      &
\multicolumn{1}{c}{Classification}             &
\multicolumn{1}{c}{$t_{\rm dSph}$}              
\\
\multicolumn{1}{c}{}                           & 
\multicolumn{1}{c}{}                           &
\multicolumn{1}{c}{}                           &
\multicolumn{1}{c}{(deg)}                      & 
\multicolumn{1}{c}{}                           &
\multicolumn{1}{c}{Formation}                  &
\multicolumn{1}{c}{($10^7 M_{\odot}$)}           &
\multicolumn{1}{c}{}                           &
\multicolumn{1}{c}{}                           &
\multicolumn{1}{c}{}                           &
\multicolumn{1}{c}{(Gyr)}                      
\\
\multicolumn{1}{l}{(1)}                        &
\multicolumn{1}{c}{(2)}                        &
\multicolumn{1}{c}{(3)}                        &
\multicolumn{1}{c}{(4)}                        &
\multicolumn{1}{c}{(5)}                        &
\multicolumn{1}{c}{(6)}                        &
\multicolumn{1}{c}{(7)}                        &
\multicolumn{1}{c}{(8)}                        &
\multicolumn{1}{c}{(9)}                        &
\multicolumn{1}{c}{(10)}                       &
\multicolumn{1}{c}{(11)}                       
\vspace*{-8pt}
\\
\\
\hline
\vspace*{-8pt}
\\
S1   &   0   & 1 & 45 & O1  &  Yes  & ...  & 0.28 & 0.73  & dSph       & 7.65 (5)  \\
S2   &   0   & 1 & 45 & O2  &  Yes  & ...  & 0.00 & 0.96  & dSph       & 3.75 (4)  \\
S3   &   0   & 1 & 45 & O3  &  No   & ...  & 0.69 & 0.40  & Non-dSph   & ...       \\
S4   &   0   & 1 & 45 & O4  &  Yes  & ...  & 0.01 & 0.97  & dSph       & 4.00 (3)  \\
S5   &   0   & 1 & 45 & O5  &  No   & ...  & 0.80 & 0.50  & Non-dSph   & ...       \\
S6   &   0   & 1 & 45 & O6  &  No   & ...  & 0.76 & 0.61  & Non-dSph   & ...       \\
S7   &   0   & 1 & 45 & O7  &  Yes  & ...  & 0.39 & 0.54  & Non-dSph   & ...       \\
\hline
\vspace*{-0.2cm}
\\
S8   &  0.8  & 1 & 45 & O1  &  No   & 0.00 & 0.12 & 0.95  & dSph       & 3.55 (2)  \\
S9   &  0.8  & 1 & 45 & O2  &  No   & 0.00 & 0.01 & 0.89  & dSph       & 2.05 (2)  \\
S10  &  0.8  & 1 & 45 & O3  &  No   & 2.34 & 0.89 & 0.62  & Non-dSph   & ...       \\
S11  &  0.8  & 1 & 45 & O4  &  No   & 0.00 & 0.03 & 0.90  & dSph       & 2.70 (2)  \\
S12  &  0.8  & 1 & 45 & O5  &  No   & 0.00 & 0.15 & 0.94  & dSph       & 7.65 (4)  \\
S13  &  0.8  & 1 & 45 & O6  &  No   & 0.00 & 0.12 & 0.94  & dSph       & 5.80 (4)  \\
S14  &  0.8  & 1 & 45 & O7  &  No   & 0.00 & 0.33 & 0.88  & dSph       & 5.60 (2)  \\
\hline
\vspace*{-0.2cm}                                           
\\
S15  &  0.5  & 1 & 45 & O1  &  No  & 0.00 & 0.02 & 0.96  & dSph        & 4.05 (2)  \\
S16  &  0.5  & 1 & 45 & O2  &  No  & 0.00 & 0.01 & 0.82  & dSph        & 2.15 (2)  \\
S17  &  0.5  & 1 & 45 & O3  &  No  & 1.04 & 1.05 & 0.53  & Non-dSph    & ...       \\
\hline
\vspace*{-0.2cm}
\\
S18  &  0.8  & 2 & 45 & O1  &  No  & 0.00 & 0.10 & 0.92 & dSph         & 3.55 (2)  \\
S19  &  0.8  & 2 & 45 & O2  &  No  & 0.00 & 0.00 & 0.86 & dSph         & 2.05 (2)  \\
S20  &  0.8  & 2 & 45 & O3  &  No  & 0.97 & 0.75 & 0.75 & Non-dSph     & ...       \\
\hline                                              
\vspace*{-0.2cm}                                           
\\
S21  &  0.8  & 1 & 0  & O1  &  No  & 0.00 & 0.08 & 0.93  & dSph        & 3.65 (2)  \\
S22  &  0.8  & 1 & 90 & O1  &  No  & 0.00 & 0.14 & 0.94  & dSph        & 3.65 (2)  \\
\vspace*{-7pt}
\\
\hline
\end{tabular}
\end{center}
\label{table:summary}
\end{table*}



\begin{figure*}
\begin{center}
\begin{tabular}{c}
  \includegraphics[scale=0.42]{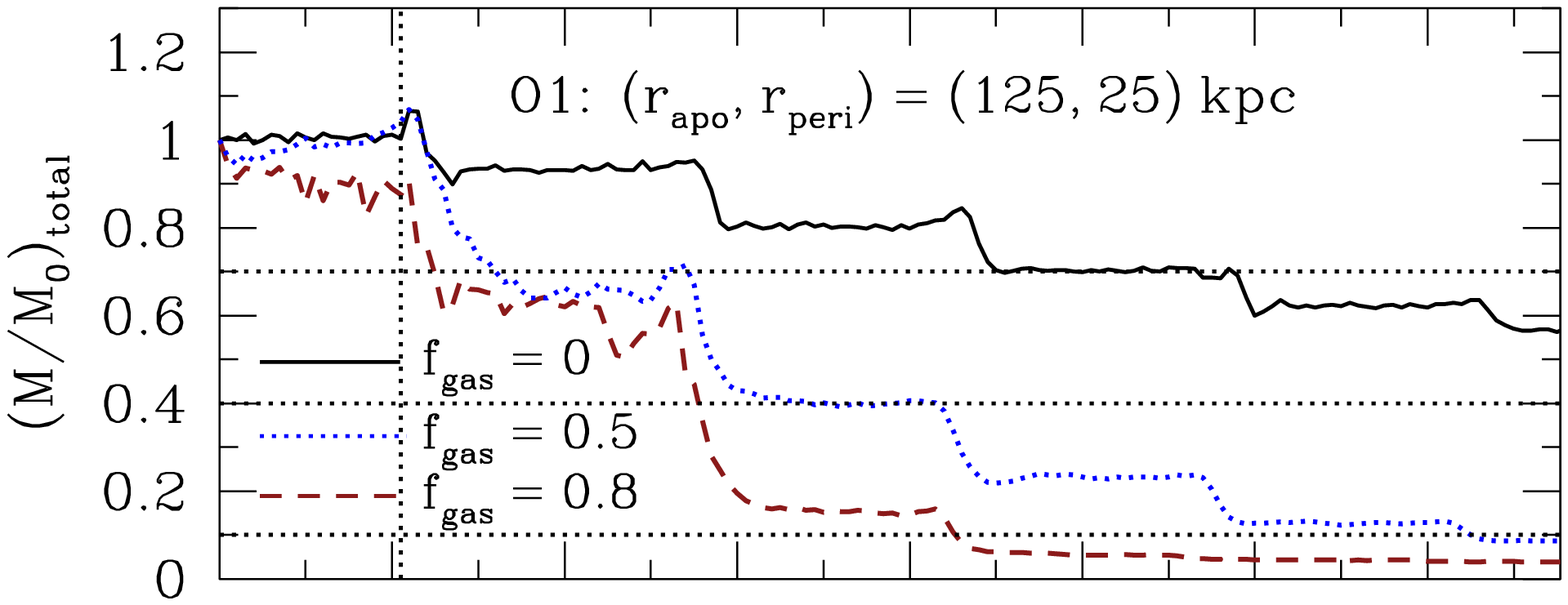}\\ 
  \includegraphics[scale=0.42]{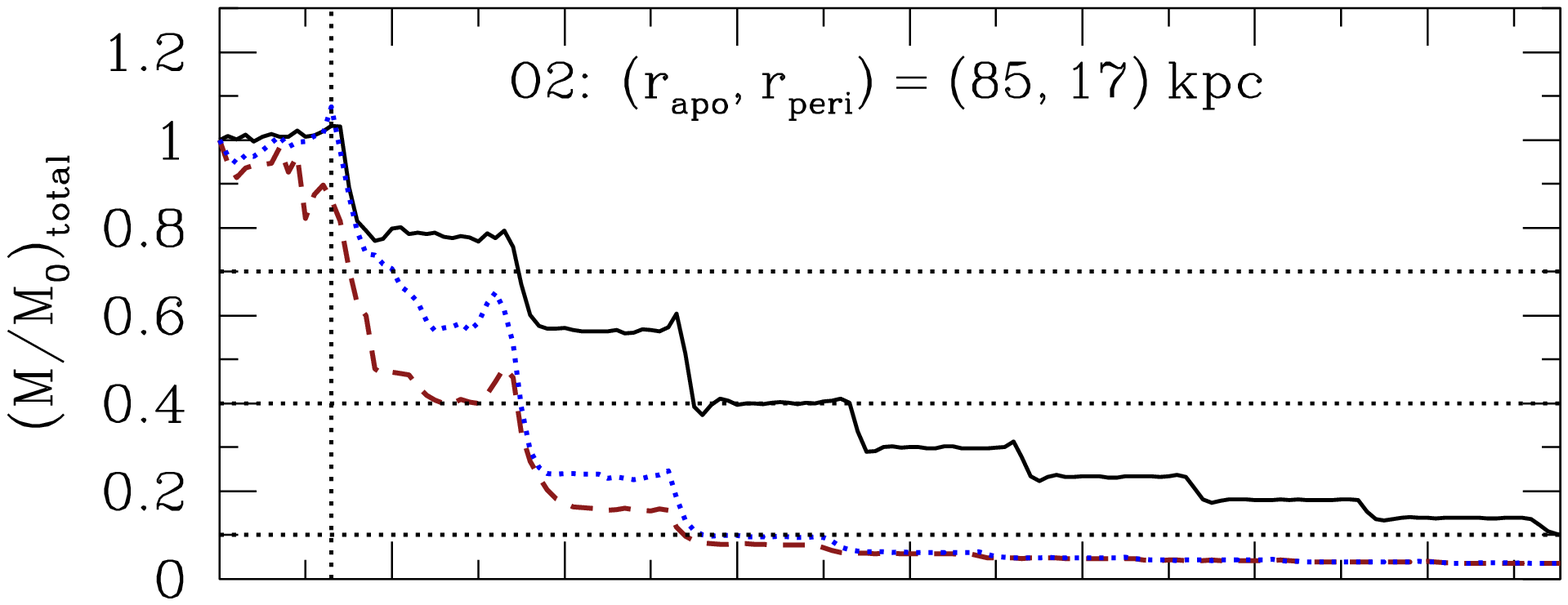}\\ 
  \includegraphics[scale=0.42]{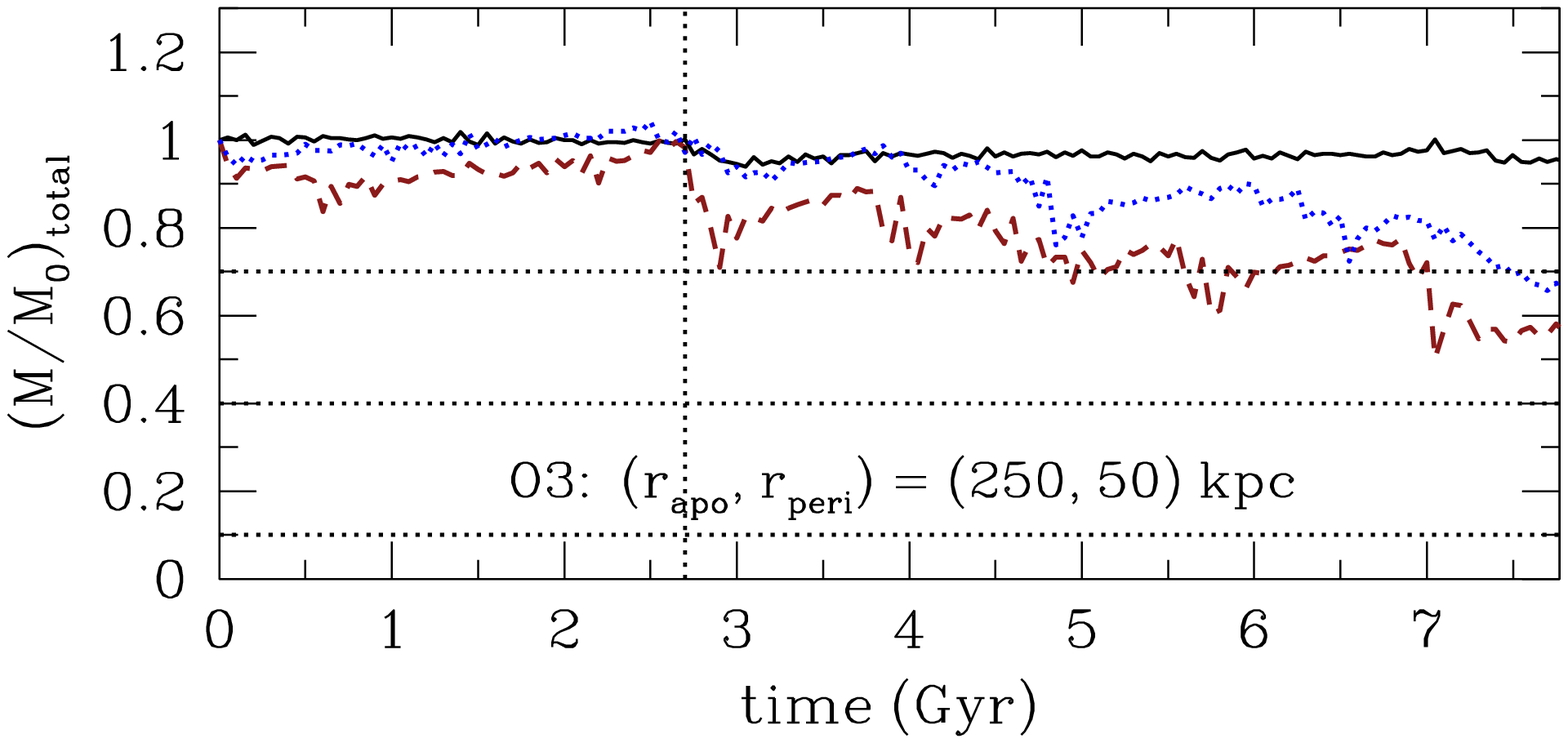}
\end{tabular}
\hspace{-0.5cm}
\begin{tabular}{c}
  \includegraphics[scale=0.42]{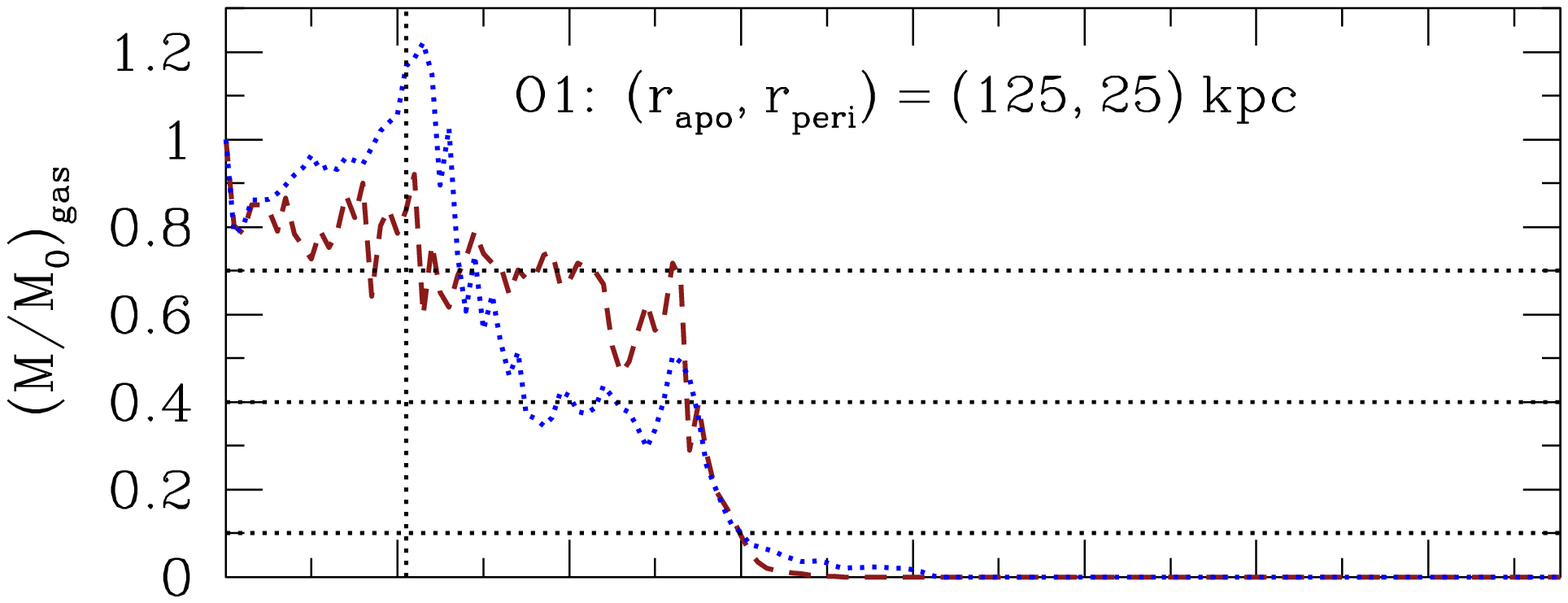}\\
  \includegraphics[scale=0.42]{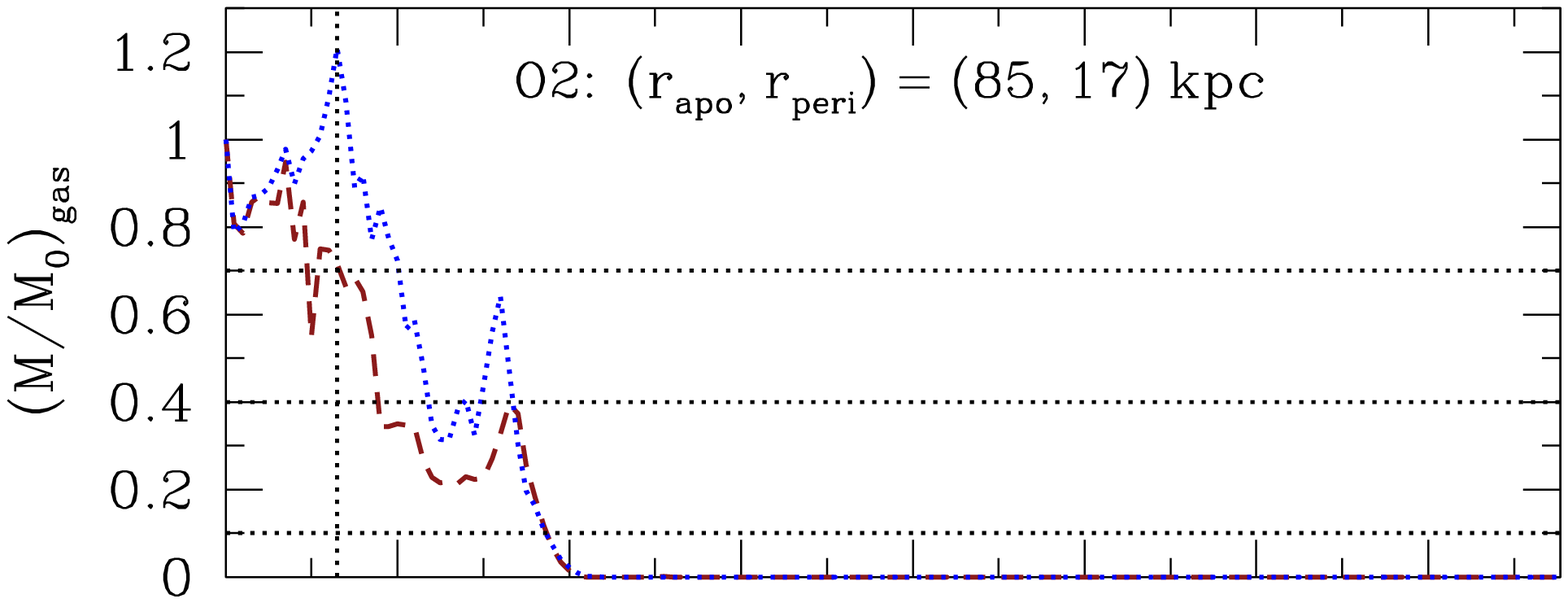}\\
  \includegraphics[scale=0.42]{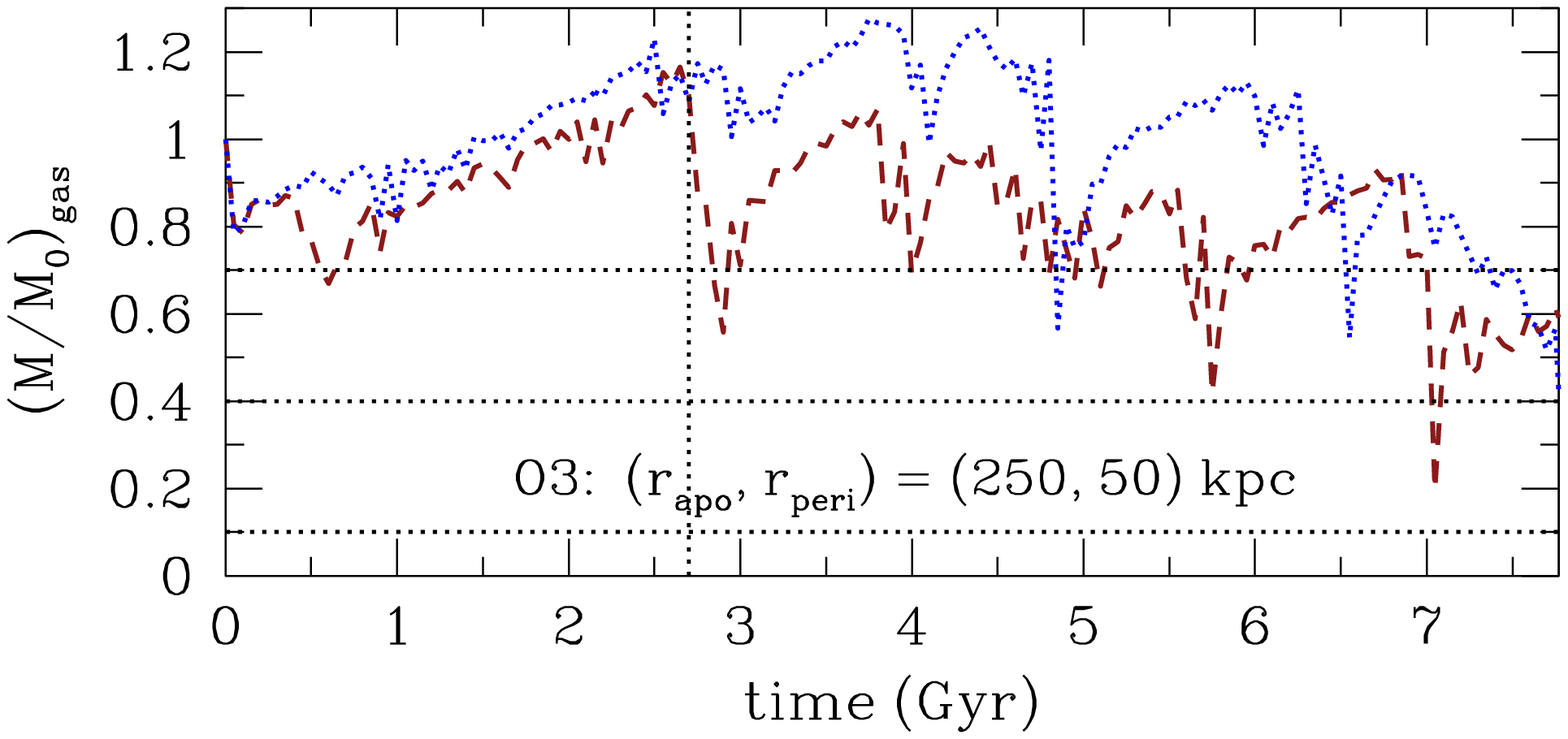}
\end{tabular}
\end{center}
\vspace{-0.5cm}
\caption{Evolution of total (left panels) and gas (right panels) mass
  of the simulated disky dwarf galaxies as a function of time. The
  results are presented for orbits O1 (upper panels), O2 (middle
  panels), and O3 (lower panels). Line types are as in
  Figure~\ref{fig1}. Total (gas) masses are computed within $0.76$~kpc
  from the center of the dwarf (see the text for details) and are
  normalized to the initial total (gas) mass enclosed within
  $0.76$~kpc, $M_0$. In all panels, horizontal lines indicate mass
  loss of $30\%$, $60\%$, and $90\%$ with respect to $M_0$. For a
  given orbit, gas-rich ($f_{\rm gas} \gtrsim 0.5$) rotationally
  supported dwarfs suffer enhanced mass loss compared to their
  collisionless ($f_{\rm gas} =0$) counterparts.
  \label{fig2}}
\end{figure*}


Figure~\ref{fig2} presents, in representative cases, the time
evolution of both total and gas mass of the dwarfs and offers insight
into these findings. In addition to illustrating the continuous
decrease of the dwarf masses by the primary tidal field, this figure
demonstrates that stripping of the total mass depends sensitively on
$f_{\rm gas}$. Indeed, while mass loss from the inner regions of the
collisionless dwarf galaxies is relatively gentle and gradual,
stripping is significantly boosted for gas-rich dwarfs. The strongest
episodes of mass loss coincide with pericentric passages, where
ram-pressure stripping is also maximal. At such times, the differences
with the collisionless simulations are remarkable and become even more
pronounced with higher gas fractions.

The rapid removal of gas via ram pressure, aided by SN explosions,
reduces swiftly the central densities of the gas-rich disky
dwarfs. Therefore, these systems should respond more impulsively to
tides and experience stronger tidal effects relative to their
collisionless counterparts. Indeed, in the impulse approximation,
tidal shocks lead to an energy increase given by $\Delta E / E \propto
R^3 / M$, where $M$ corresponds to the total dwarf mass within a
characteristic radius $R$ (see \citealt{Kazantzidis_etal11} for a
detailed discussion). Owing to ram-pressure stripping, at a given
distance $R$ from the center of our dwarfs, increasing gas fractions
correspond to smaller $M$, and thus to larger $\Delta E / E$; hence,
the gas-rich dwarfs suffer more effective tidal shocks, which explains
both their enhanced mass loss and augmented morphological
transformation into dSph-like systems.  Taking into account adiabatic
corrections \citep[e.g.,][]{Gnedin_Ostriker99} strengthens this
conclusion. Indeed, for less concentrated mass distributions,
adiabatic corrections result in even larger $\Delta E / E$ compared to
those predicted by the impulse approximation.

Apart from amplifying ram-pressure stripping, SN winds play another
crucial role in causing the gas-rich dwarfs to exhibit a more
impulsive response to tides. This role is revealed in
Figure~\ref{fig2}, which shows that the $f_{\rm gas}=0.8$ dwarf
galaxies reach their first pericenter, thus in the absence of any
tidal shocks, already with reduced total central mass. Such a decrease
is due to rapid gas outflows at sub-kiloparsec scales, driven by
multiple SN explosions, which induce strong and repeated fluctuations
in the dwarf gravitational potential. Such fluctuations result in
energy transfer to the collisionless components (DM and stars),
significantly lowering the total central densities of dwarf galaxies
\citep[e.g.,][]{Governato_etal10,Pontzen_Governato12,Teyssier_etal13,Shen_etal14,Onorbe_etal15,Read_etal16}.
Due to the increased SF, the overall effect of SN winds is expected to
be stronger for higher gas fractions. In conjunction with the impact
of ram pressure, this naturally explains the trend of augmented
transformation efficiency with larger $f_{\rm gas}$.

\section{Discussion}
\label{sec:discussion}

The present study is the first to elucidate the combined effects of
radiative cooling, ram-pressure stripping, SF (based on a high gas
density threshold), SN feedback, and a cosmic UV background on the
tidal stirring mechanism for the origin of dSphs. We have shown that,
regardless of orbit inside the primary galaxy, gas-rich ($f_{\rm gas}
\gtrsim 0.5$) rotationally supported dwarfs exhibit a substantially
enhanced likelihood and efficiency of transformation into dSphs
relative to their collisionless ($f_{\rm gas} = 0$) counterparts
(Columns 7–11; Table~\ref{table:summary}). Such large values of
$f_{\rm gas}$ are akin to those of observed dIrrs
\citep[e.g.,][]{McConnachie12,Oh_etal15}, suggesting that the
morphology-density relation \citep[e.g.,][]{Mateo98}, an essential
constraint that any theoretical model for the LG must satisfy, may
naturally arise in the context of the tidal stirring model.

We gauged the effect of ram pressure on our results by repeating
simulation S8 without the hot gas halo in the host. In the absence of
ram pressure, the gas was not entirely stripped from the dwarf ($\sim
40\%$ of the original gas content still remained bound). With the
obvious caveat that we only explored a single orbit, this experiment
suggests that ram pressure is required for the formation of bona fide
dSphs via the tidal stirring of gas-rich rotationally supported
dwarfs. Our finding that gravitational tides alone, although aided by
SN feedback in our case, cannot entirely remove the gas components
from dwarfs, is in agreement with \citet{Mayer_etal06}. 

In addition, we ascertained how the transformation process depends on
SN feedback by resimulating experiment S8 without this effect. The
absence of SN explosions resulted in a notably decreased efficiency of
dSph formation, with the timescales of ram-pressure stripping becoming
significantly longer (see, also,
\citealt{Mayer_etal06}). Specifically, complete gas removal now takes
place at $5.85$~Gyr (and after four pericentric passages), compared to
$3.55$~Gyr (and after two pericentric passages) in the original
simulation S8. This highlights the vital role of SN winds in enabling
rapid ram-pressure stripping by making cold gas more loosely bound and
by carving holes in the ISM with expanding bubbles that increase
ablation effects \citep[e.g.,][]{Murakami_Babul99,Marcolini_etal04}.

Although the exclusion of SN feedback did not affect the overall
likelihood of transformation, as a bona fide dSph still formed, it did
influence the degree of transformation. Indeed, the final values of
$V_{\rm rot}/\sigma_{\ast}$ and $c/a$ were equal to $0.25$ and $0.81$,
respectively (compared to $V_{\rm rot}/\sigma_{\ast} = 0.12$ and
$c/a=0.95$, when SN feedback was included; see
Table~\ref{table:summary}). As discussed in Section~\ref{sec:results},
the less complete transformation in this case is attributed to both
the less efficient ram-pressure stripping and the absence of gas
outflows from SN winds, causing a more adiabatic response of the disky
dwarfs to tidal effects.

Previous work on tidal stirring that did not include SN explosions
indicated that the timescales of ram-pressure stripping, and
consequently those of dSph formation, were sensitive to the presence
of a cosmic ionizing background \citep[e.g.,][]{Mayer_etal06}. We
examined this assertion by repeating simulation S8 without the cosmic
UV and found that its absence had virtually no effect on either the
timescales of ram-pressure stripping or on the efficiency, likelihood,
and degree of transformation. Similar conclusions were reached in the
experiments with a stronger cosmic UV background (S18--S20;
Table~\ref{table:summary}). These results again underscore the
fundamental importance of SN feedback as the primary heating mechanism
for the gas and the main effect that facilitates ram-pressure
stripping.

We also investigated whether the alignment between the internal and
orbital angular momentum of the dwarfs can affect their
transformation. To this end, we repeated simulations S1 and S8,
adopting a {\it mildly} retrograde alignment of $i=-45\degrees$. The
analysis of these experiments showed that the collisionless disky
dwarf does not transform into a dSph (see, also,
\citealt{Lokas_etal15}), highlighting the importance of coupling
between orbital and internal motions; however, its gas-rich
counterpart does experience transformation, although with
significantly decreased efficiency ($t_{\rm dSph}= 6.1$~Gyr, as
opposed to $t_{\rm dSph}= 3.55$~Gyr in the prograde simulation;
Table~\ref{table:summary}). In addition, the ram-pressure stripping
timescales were longer and the degree of transformation was
substantially weaker in the retrograde case. Mildly retrograde
alignments thus seem to support the general picture presented in this
study.

Earlier tidal stirring studies established a strong association
between the development of long-lived bars in the disks of the
progenitor disky dwarfs and dSph formation
\citep[e.g.,][]{Mayer_etal01a,Mayer_etal01b,
  Klimentowski_etal09,Kazantzidis_etal11,Kazantzidis_etal13}. Nonetheless,
with the prominent exception of the LMC, the number of irrefutable
bar-like distortions observed in LG dwarfs is extremely low (see,
however, \citealt{Lokas_etal12}), challenging the overall predictive
power of the tidal stirring model.

Interestingly, our results indicate that bar formation is not a {\it
  necessary} condition for the transformation of rotationally
supported gas-rich dwarfs via tidal stirring (Column 6;
Table~\ref{table:summary}), alleviating the tension between
observations and theoretical predictions. In these cases, the lower
stellar mass fractions, which correspond to a lower self gravity of
the tidally perturbed stellar disk, prevent the development of a bar,
and dSph-like systems are produced only via tidal heating.

Lastly, our findings demonstrate a correlation between the amount of
residual gas in the dwarfs and their degree of transformation into
dSphs (Table~\ref{table:summary}). Specifically, total gas stripping
is required for the most complete transformations, while the presence
of gas is associated with objects that maintain, at least partially,
the properties of their disky progenitors (S10, S17, S20). The latter
finding is particularly relevant to the ``transition-type'' dwarfs
(e.g., Phoenix, Pegasus, and LGS3), which share properties with both
dSphs and dIrrs \citep{Grcevich_Putman09}. One possibility for their
origin is that their progenitors were disky satellites that were only
recently accreted by their hosts on very wide, radial orbits.  These
transition-type dwarfs could therefore still be in the process of
being transformed into dSphs by tidal stirring, having concluded one
pericentric passage and having some of their gas removed via ram
pressure. In fact, ram-pressure stripping has been proposed to explain
the presence of $H_{\scriptsize I}$ clouds in the vicinity of Phoenix
\citep{Gallart_etal01}. In the aforementioned scenario,
transition-type dwarfs should possess appreciable stellar rotation,
having been only partially transformed. Although recent studies have
reported significant rotation in both Pegasus \citep{Wheeler_etal17}
and Phoenix \citep{Kacharov_etal17} with $V_{\rm rot}/\sigma_{\ast}
\approx 1$, irrefutable evidence of substantial rotation in these
systems would be required to validate our prediction.

\acknowledgments

We acknowledge stimulating discussions with Thanos Anestopoulos,
Giuseppina Battaglia, and David Weinberg. S.K. was supported by the
Pauli Center for Theoretical Studies at the University of Z\"urich
during the final stages of this work.  The numerical simulations were
performed at JPL.

\end{document}